\documentclass[prl,showpacs,twocolumn]{revtex4}
\usepackage{graphicx}

\begin{document}

\title{Multiple soft-mode vibrations of lead zirconate }

\author{J. Hlinka }
\email{hlinka@fzu.cz}
\author{T. Ostapchuk}
\author{E. Buixaderas}
\author{C. Kadlec}
\author{P. Kuzel}
\author{I. Gregora}
\author{J. Kroupa}
\author{M. Savinov}
\author{J. Drahokoupil}
 \affiliation{
Institute of Physics, Academy of Sciences of the Czech Republic\\%
Na Slovance 2, 182 21 Prague 8, Czech Republic}
\author{J. Dec}
 \affiliation{Institute of Materials Science, University of Silesia, Bankowa 12,
PL-40-007 Katowice, Poland}
\date{\today}

\begin{abstract}

 Polarized Raman, IR and time-domain THz  spectroscopy of orthorhombic lead zirconate single crystals yielded a
 comprehensive picture of  temperature-dependent quasiharmonic frequencies of its low-frequency phonon modes.
 It is argued that these modes primarily involve vibration of Pb and/or oxygen octahedra librations and their relation to particular phonon modes of the parent cubic phase is proposed. Counts of the observed IR and Raman active modes  belonging to  distinct irreducible representations agree quite well with group-theory predictions. The most remarkable finding is the considerably enhanced frequency renormalization of the $y$-polarized polar modes, resulting in a
 pronounced  low temperature dielectric anisotropy. Results are discussed in terms of contemporary phenomenological theory of antiferroelectricity.
\end{abstract}


\maketitle

Although the ferroelectric and antiferroelectric materials have a lot in common,
the latter have been much less investigated. An obvious reason is the absence of
the direct linear coupling of the antiferroelectric (AF) order parameter
to the macroscopic electric field.

At the same time, a {\it nonlinear} coupling to the macroscopic electric field is still present.
Therefore, AF materials
 actually do provide interesting functionalities, as well.
 In fact,  the AF oxides are promising
materials for high-energy storage capacitors, high-strain actuators and perhaps even for electrocaloric refrigerators\cite{Rabe13,LiuH11,Misc06}.
The interest in the improvement of our understanding of AF oxides has been expressed recently\cite{Rabe13,Taga13,LiuH11,Reye13}.

\begin{figure}[h]
\includegraphics[width=65mm]{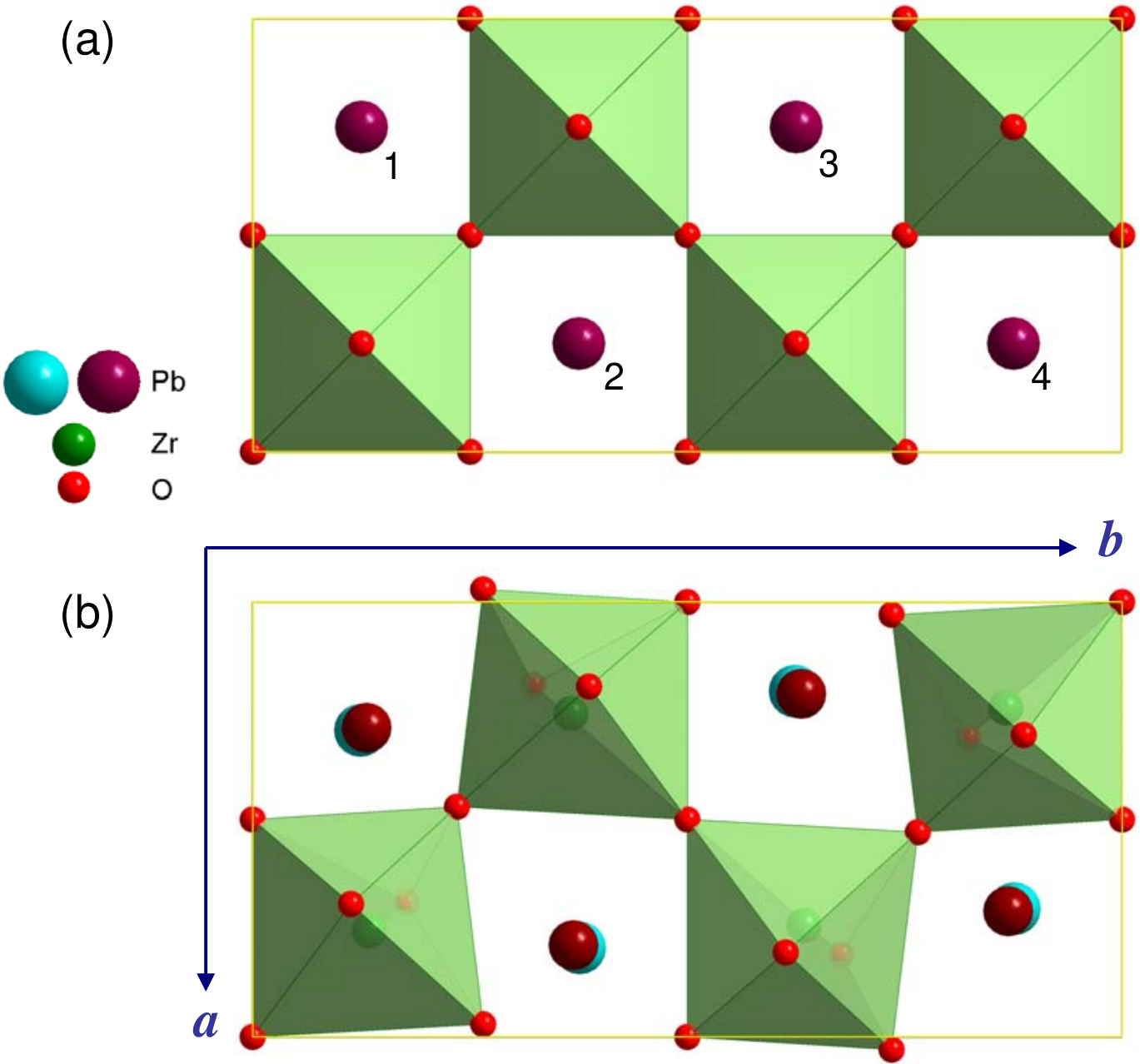}
 \caption{(Color online)
 Schematic illustration of the crystal structure of PbZrO$_3$ in its  (a)  high-temperature cubic phase and (b) in its low-temperature orthorhombic phase. Orthorhombic $Pbam$ elementary unit cell is projected along
 its $z\parallel {\bf c}$ axis,  the  ${\bf a}= {(1,-1,0)_{pc}/\sqrt{2}}$  and  ${\bf b}= {(1,1,0)_{pc}\sqrt{2}}$  lattice vectors  defining the $x$ and $y$ axes are indicated in the figure.
 Note that the $x$-components of the AF displacements of the Pb ions 1 and 2 are opposite to those of 3 and 4. This displacement pattern forms a $\Sigma_3$ symmetry mode associated with ${\bf Q}_{\Sigma}={\bf b^*}$ propagating vector.
  } \label{fig61}
\end{figure}

 Lead zirconate, PbZrO$_3$,  is the best known example of an
AF oxide - it is an end-member of technologically
relevant solid solutions with PbTiO$_3$ (piezoelectric PZT's)\cite{Taga13,Rabe13,LiuH11,Lines,Shir50,Bhal00}. The parent paraelectric phase is a simple cubic perovskite with a 5-atom unit cell
($Pm\bar{3}m$, Z=1).
Below the AF phase transition ($T_{\rm C} \sim 500$\,K), it goes over into an
orthorhombic $Pbam$
(Z=8) structure\cite{Fuji84,Coch68}.
The space-group symmetry change can be well understood\cite{Rabe13} as a result of the
condensation of two
order parameters\cite{Fuji68, Rabe13,Taga13,Wagh97}.  One of them is a polarization wave of  a propagation vector ${\bf Q}_{\Sigma}=(0.25,0.25,0)_{\rm pc}$,  the other order parameter is a ${\bf Q}_{R}=(0.5,0.5,0.5)_{\rm pc}$ oxygen octahedra tilt mode (here pc stands for pseudocubic lattice, see Figs.\,1-2).

 Superpositions of ${\bf Q}_{\Sigma}$, ${\bf Q}_{R}$ include also  $\Gamma, X, M$ and ${\bf Q}_{S}=(0.25,0.25,0.5)_{\rm pc}$ cubic-phase Brillouin zone points. All of these points become Brillouin zone centers in the $Pbam$ phase (see Fig.\,2).
 Nevertheless, recent inelastic X-ray scattering experiments\cite{Taga13} have clearly demonstrated that the critical scattering occurs only in the vicinity of the $\Gamma$-point. Based on this experimental result, it was proposed that the AF phase transition is driven by a {\it single mode}, the $\Gamma$-point ferroelectric soft mode\cite{Taga13}.  Within this model, the condensation of the ${\bf Q}_{\Sigma}$-point mode can be ascribed to the flexoelectric coupling with the ferroelectric mode, and the condensation of the ${\bf Q}_{R}$-point  mode can be explained  as due to a biquadratic coupling with the ${\bf Q}_{\Sigma}$ mode (i.e. through the Holakovsky\cite{Hola73} triggering mechanism)\cite{Taga13}.

\begin{figure}[h]
\includegraphics[width=60mm]{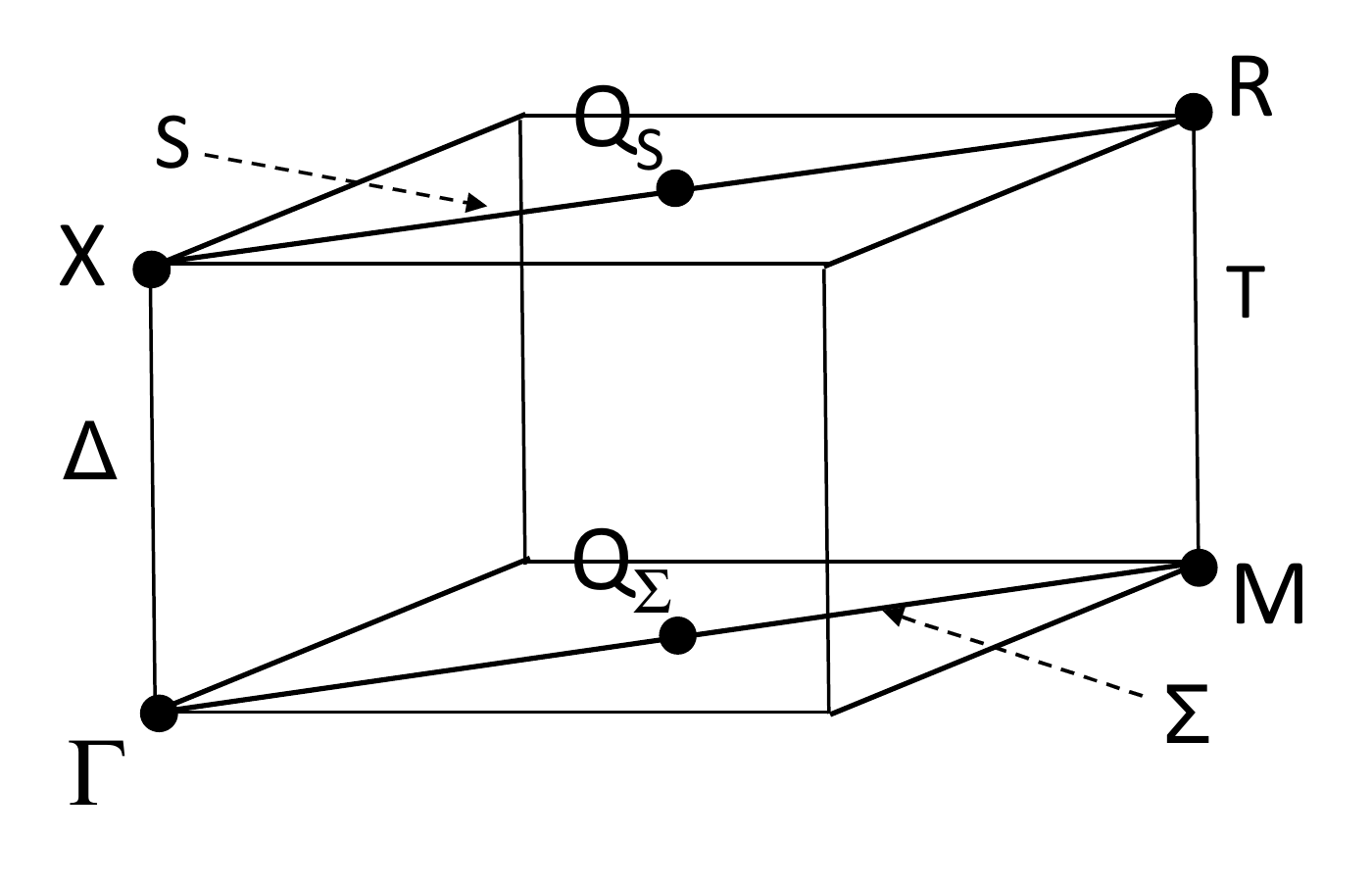}
 \caption{
 Part of the Brillouin zone of the cubic PbZrO$_3$ crystal with indicated  $\Gamma, X, R, M, {\bf Q}_{\Sigma}$ and ${\bf Q}_{S}$ Brillouin zone points.
  } \label{fig62}
\end{figure}

 It should be noted, however, that the earlier spectroscopic studies\cite{Doba01,Role89,Osta01} indicated the existence of additional lattice modes with temperature dependent frequencies.
  The principal aim of this work is to provide a systematic overview of the temperature dependence of the low frequency phonon modes of the AF PbZrO$_3$ by means of polarized IR and Raman spectroscopic study of single domain specimens.
  The results obtained  testify the existence of {\it multiple soft modes} of different symmetry in orthorhombic PbZrO$_3$. Consequently, the simple
  scenario with a single driving lattice mode as proposed in Ref.\,\onlinecite{Taga13} has to be modified.

 Before discussing the novel experimental results, let us note that the state-of-art density functional theory  calculations have clearly demonstrated that  the parent cubic structure of PbZrO$_3$ is unstable at low temperatures with respect to the Pb ion off-centering as well as concerted oxygen octahedra tilts\cite{Ghos99,Reye13,Wagh97,Cock00,Leun02,Leun03}. These calculations show a system of unstable branches, dominated by Pb-O vibration, and including  $\Gamma_{15}, M_{5}', M_{2}', X_{5}', X_{2}'$ and $R_{15}$ phonon modes,  as well as a few unstable branches connecting the rigid-body oxygen-octahedra tilt modes $ M_{3}, X_{5}$ and $R_{25}$ (throughout the paper, we are using the labels of Ref.\,\onlinecite{Ghos99}.)

 The PbZrO$_3$ phonon dispersions curves  have not yet been determined experimentally. Nevertheless,  since the cubic-phase  properties of PbTiO$_3$-PbZrO$_3$ solid solutions (PZT) are expected to vary  smoothly with the PbTiO$_3$ concentration, the cubic-phase phonon frequencies of Pb-dominated modes  can be  estimated from the previous inelastic neutron and X-ray scattering studies of PbTiO$_3$ and PZT. For example,
 the frequencies of the $M_{5}', M_{2}', X_{5}', X_{2}'$ and $R_{15}$ Pb-dominated modes can be extrapolated from the  measurements of the  PbTiO$_3$\cite{Shir70,Tome12,Kemp06} and the  PZT single crystals\cite{Hlin11};  the  acoustic mode dispersion can be estimated  from PbZrO$_3$ measurements of Refs.\,\onlinecite{Taga13,KoJH13}, and the zone-center mode frequency can be estimated from the dielectric measurements\cite{Osta01, Taga13, Buix11}.
 The lowest frequency phonon dispersion curves of the cubic PbZrO$_3$ obtained in this way (near the phase transition point) are  traced    in Fig.\,3. Let us note that almost all the estimated Pb-dominated mode frequencies fall below about  100\,cm$^{-1}$,  while those of $ M_{3}, X_{5}$ and $R_{25}$  oxygen octahedra tilt modes (not shown in Fig\,3.) are expected to lie {\it above} 100\,cm$^{-1}$.

\begin{figure}[h]
 \includegraphics[width=80mm]{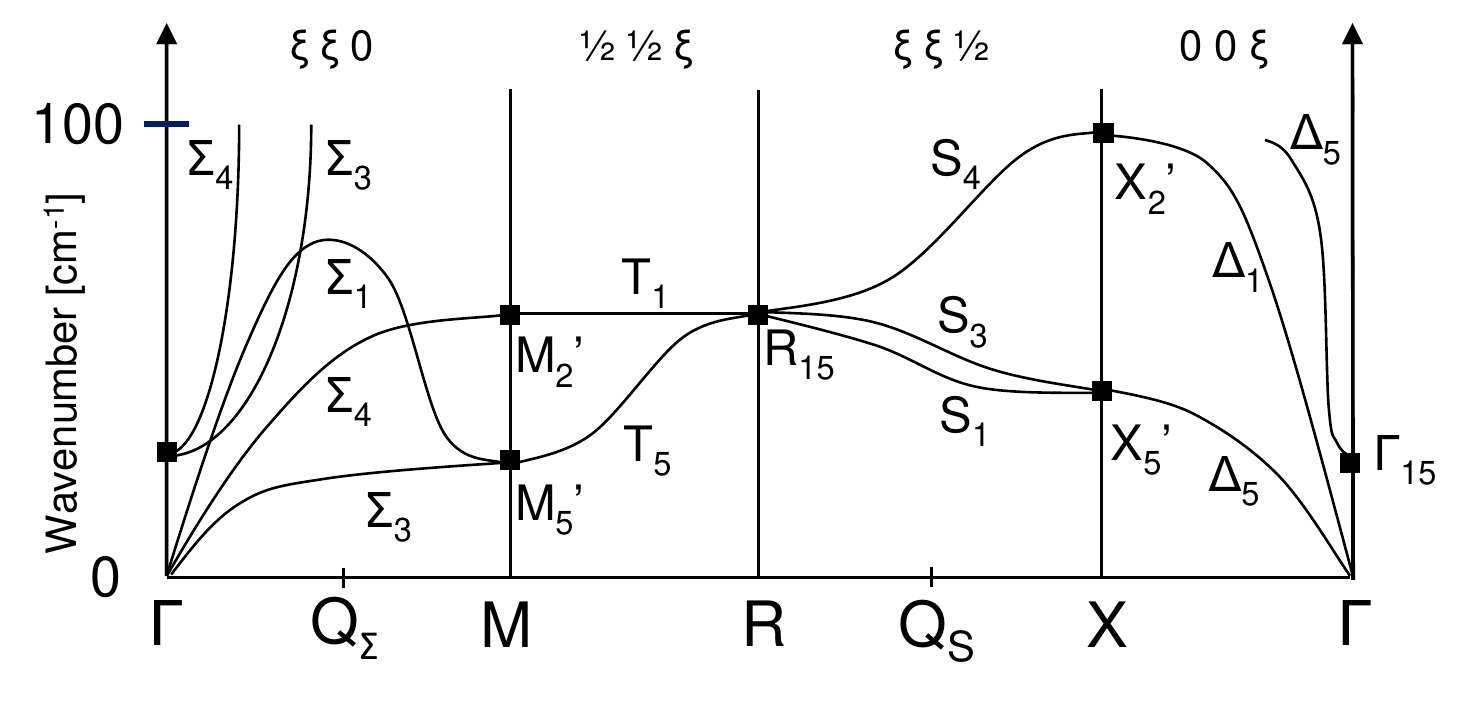}
 \caption{
   Low frequency phonon branches of cubic PbZrO$_3$. Phonon frequencies at the $\Gamma,  M, R $ and $X $ points are estimated from the available PbTiO$_3$ and PbTiO$_3$-PbZrO$_3$ spectroscopic data. Indices of the symmetry labels denoting phonon branches are those of Ref.\,\onlinecite{Cowl}.
  } \label{fig63}
\end{figure}


The cubic-orthorhombic transition is of the first order, but the PbZrO$_3$ orthorhombic structure can be well understood as due to a small structural distortion of the cubic one. Thus, in the limit of the vanishing distortion, the $16A_g + 16 B_{1g}(xy) + 14 B_{2g}(xz) + 14 B_{3g}(yz) + 12 A_{u} + 12 B_{1u}(z) + 17 B_{2u}(y) + 17 B_{3u}(x)$ $\Gamma$-point modes of the $Pbam$ orthorhombic structure   transform also as $\Gamma, X, R, M, {\bf Q}_{\Sigma}$ or ${\bf Q}_{S}$-point modes of the parent cubic phase. Correlation between irreducible representation of the actual and parent symmetry group for the Pb ion vibration modes  is shown in Table I. In fact, irreducible representations listed in Table I match well those realized in Fig.\,3. Therefore, about 24 optic Pb ion   modes are expected  in the AF phase within the $0-100$\,cm$^{-1}$ frequency range.

\begin{table}
\caption{Correlation between $D_{2h}$ irreducible representations of $Pbam$ Pb-ion zone center vibrations (top row of the table) and their counterparts in the parent cubic phase. $\Sigma_i$ and $S_i$ stands for  modes associated with ${\bf Q}_{\Sigma}$ and ${\bf Q}_{S}$ wave vectors, respectively. Other labels are as those of Ref.\,\onlinecite{Ghos99}.}\label{T1}
\begin{tabular}{l|cccccccc}
\hline
   &   A$_g$ & B$_{1g}$  &  B$_{2g}$ & B$_{3g}$ & A$_u$ & B$_{1u}$ &B$_{2u}$  &B$_{3u}$ \\
\hline
 $\Gamma$  &    &   &   &   &  & $\Gamma_{15}$ & $\Gamma_{15}$  & $\Gamma_{15}$ \\
  X  &    &   &   &   &  & X$_{2}'$ & X$_{5}'$  & X$_{5}'$ \\
    M  &  M$_{5}'$  &  M$_{5}'$   &  &  M$_{2}'$  &  &  &    &   \\
    R  &  R$_{15}$  &  R$_{15}$   &  &  R$_{15}$  &  &  &    &   \\
  $\Sigma$  &  $\Sigma_{3}$  & $\Sigma_{1}$   & $\Sigma_{4}$ &  &  $\Sigma_{4}$ &   &  $\Sigma_{3}$  & $\Sigma_{1}$    \\
S   &  S$_{3}$  & S$_{1}$   & S$_{4}$ &   & S$_{4}$ &   & S$_{3}$  & S$_{1}$    \\
\hline \hline
sum  &  4&  4&  2  & 2  & (2) & 2 & 4  & 4    \\
\end{tabular}
\end{table}


The flux-grown single crystal platelets used in the present experiments, with either out-of-plane or in-plane $c$-axis\cite{Dec89},  were detwinned using the method of Ref.\,\onlinecite{Dec89}.
 Raman data were collected using a Renishaw microscope spectrometer operated with a 514\,nm laser and a low-frequency edge filter,  like e.g. in Refs.\,\onlinecite{Raman1,Raman2}. IR reflectivity and time-domain THz transmission data were collected using a Fourier-transform Bruker spectrometer and a laboratory built system based on Ti-Sapphire laser, respectively, and then fitted simultaneously to obtain the consistent complex dielectric and conductivity spectra in the 10-800\,cm$^{-1}$ range (the same setup and experimental procedure as e.g. in Refs.\,\onlinecite{Hlin08,DielExp}).

 The typical low-frequency, low-temperature Raman spectra  are shown in Fig.\,4. As indicated in the figure, the light was polarized along the $Pbam$ crystallographic axes so that the assignment of the observed modes to the relevant irreducible representations was rather straightforward (unlike in Ref.\,\cite{Role89}). Similarly, the real part of the conductivity spectra allows probing  the $B_{1u}(z)$, $ B_{2u}(y)$ and $B_{3u}(x)$ modes independently (see Fig.\,5.).
 Overall, the numbers of the modes observed in the $0-100$\,cm$^{-1}$ frequency range corresponds quite well to the list given in Table I.

\begin{figure}
\includegraphics[width=70mm]{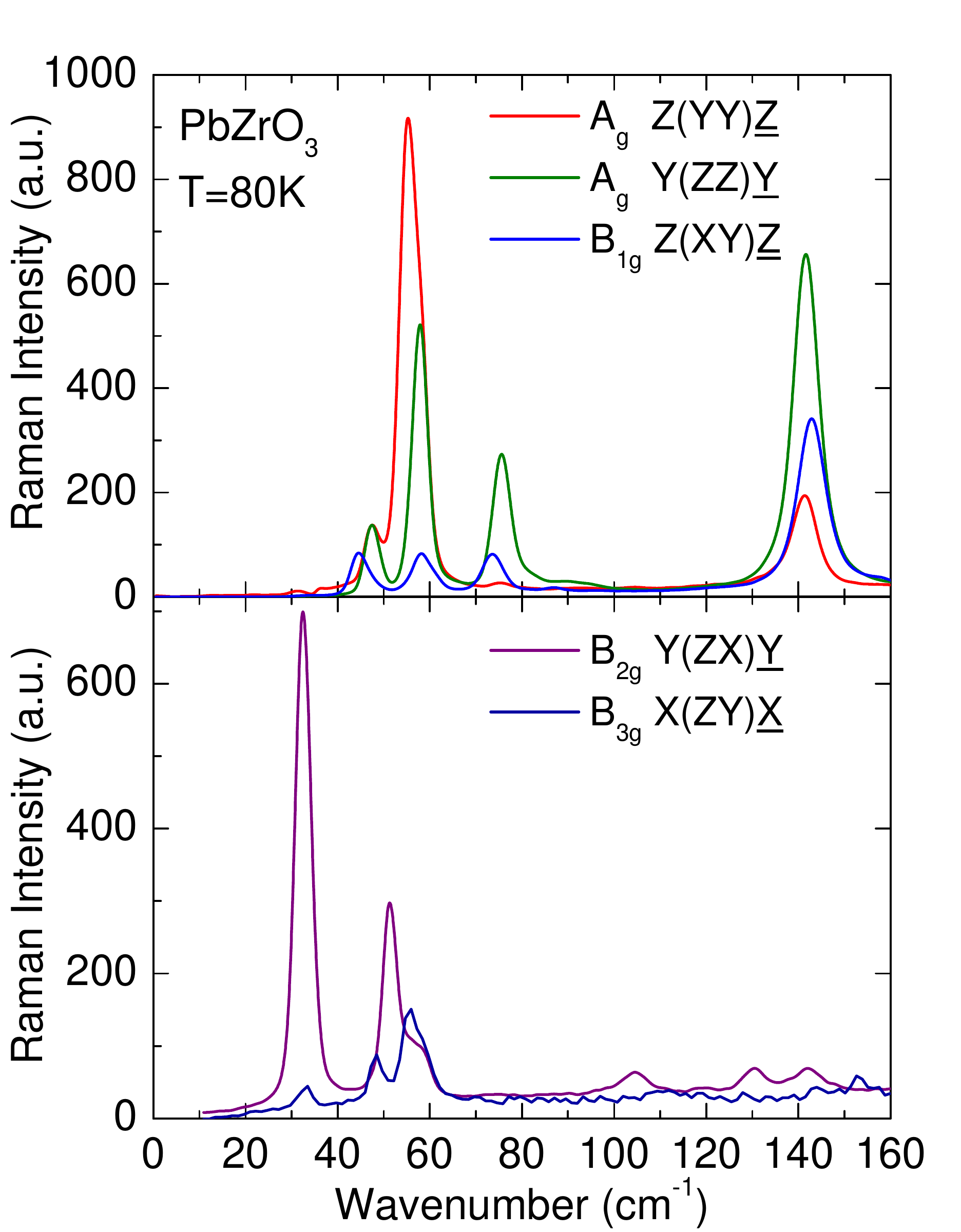}
 \caption{(Color online) Typical low-frequency polarized Raman spectra of PbZrO$_3$ single crystal at 80\,K. The scattering geometry is indicated by the usual Porto notation, where X,Y and Z are oriented along the crystallographic axes of the orthorhombic $Pbam$ structure.
  } \label{fig64}
\end{figure}

\begin{figure}[h]
\includegraphics[width=65mm]{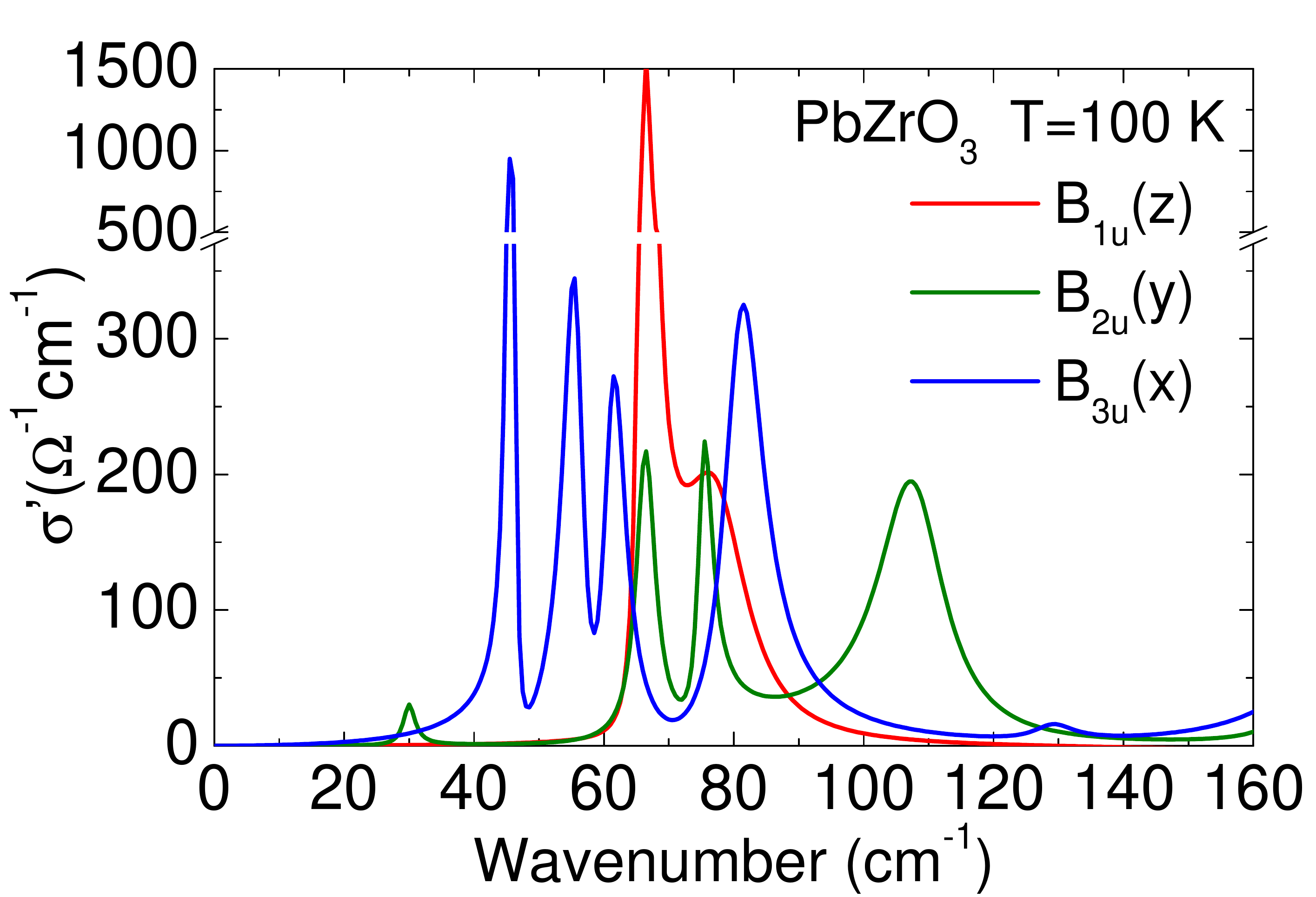}
 \caption{(Color online)
Real part of the low-frequency conductivity spectra of PbZrO$_3$, as obtained from a combined fit to spectra of IR reflectivity and time-domain THz spectroscopy, showing peaks at the transverse optic mode frequencies of $B_{1u}(z)$, $ B_{2u}(y)$ and $B_{3u}(x)$ modes, where $x,y$ and $z$ are oriented along the crystallographic axes of the orthorhombic $Pbam$ structure.
  } \label{fig65}
\end{figure}

\begin{figure}
\includegraphics[width=85mm]{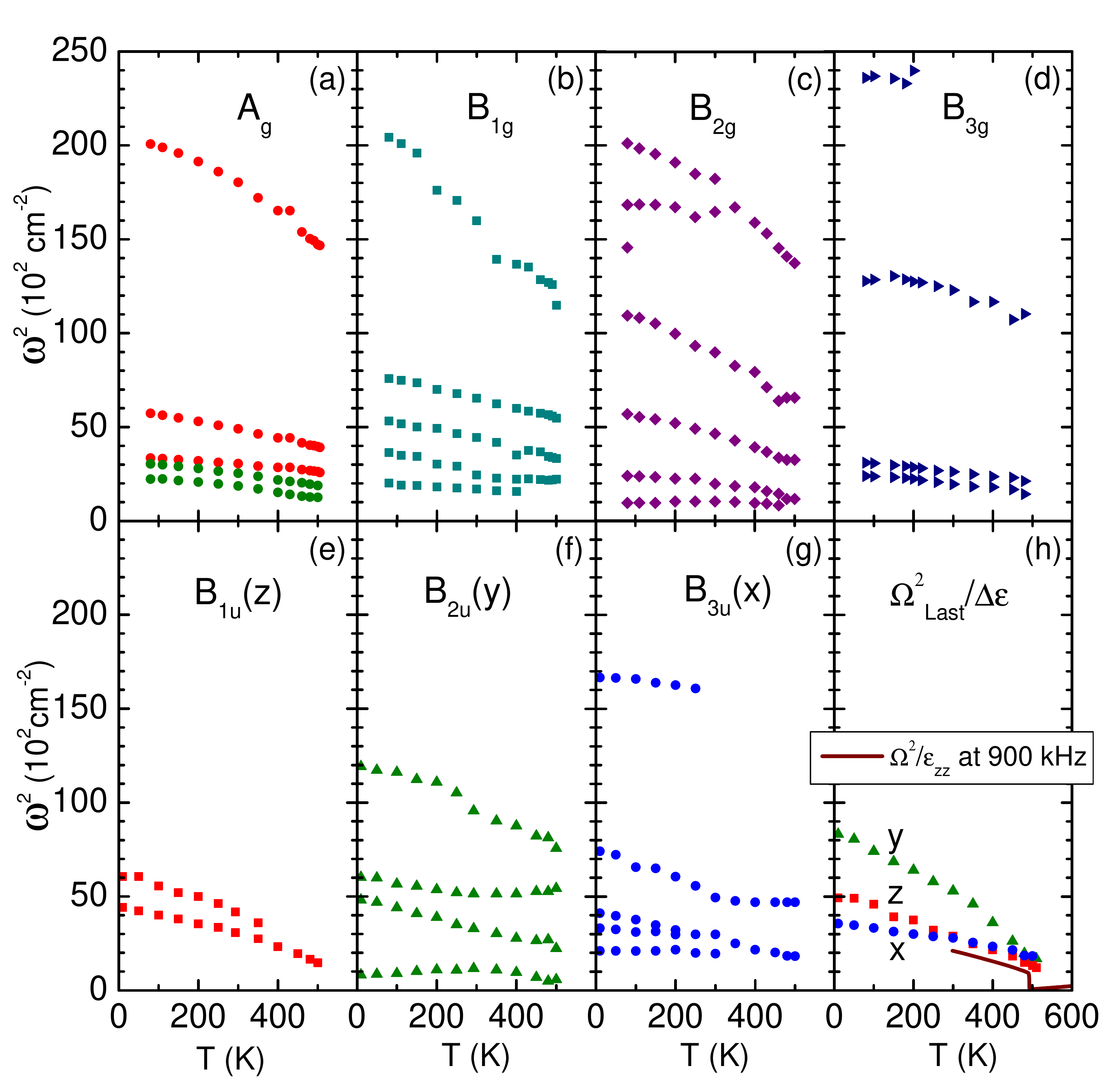}
 \caption{(Color online)
 Temperature dependence of squared frequencies of Raman active (a)-(d) and IR active (e)-(g) modes  of antiferroelectric PbZrO$_3$.
 Panel (h) shows inverse static permittivity extrapolated from the joint fit to THz and IR spectroscopy spectra, multiplied by the
 square of the plasma frequency of the Last mode ($\Omega_{\rm Last}=618$\,cm$^{-1}$). These quantities provide estimate of an effective squared oscillator frequencies of the bare Last mode components in the AF phase. The  $c$-axis inverse static permittivity measured at 900\,kHz is also shown there.
  } \label{6a7}
\end{figure}

 Phonon frequencies up to about 150\,cm$^{-1}$ (from fits using damped harmonic oscillator response functions) as a function of temperature are
 shown in Fig.\,\ref{6a7}. The lowest frequency $ B_{2u}, B_{3u}, B_{1g}$ and $B_{3u}$ modes can be assigned to the $\Sigma_3,\Sigma_1,\Sigma_1$  and $\Sigma_4$ acoustic modes folded from  $({\bf Q}_{\Sigma})$. All other modes of Fig.\,6 reveal a considerable {\it frequency increase upon cooling} (both the modes in the 0-100\,cm$^{-1}$ range as well as the modes in the 100-150\,cm$^{-1}$ frequency range).

 How can this be understood?
The temperature dependence of the fully symmetric mode ($A_g$), corresponding to the order parameter, follows naturally from the simplest Landau-type theory. The temperature dependence of the $B_{1u}$,  $B_{2u}$ and  $B_{3u}$ components of the Last-type $\Gamma_{15}$ mode could be explained  e.g. by a positive biquadratic coupling to the primary order parameter\cite{Taga13}.
What is the reason for the strikingly similar temperature dependence of so many {\it other} phonon frequencies  below 100\,cm$^{-1}$? Our understanding is  that  PbZrO$_3$ has a {\it soft-branch} driven phase transition, rather than  a soft-mode one. In other words, the observations (i) indicate a small dispersion of Pb-based phonon branches, and (ii) suggest that the stabilizing anharmonic potential has a dominantly local character, as assumed in  simple effective Hamiltonian models\cite{Wagh97,Cock00}, so that it makes both the $\Sigma_3$ and $S_3$ branches temperature dependent.  Soft phonon branches are known e.g. from  incommensurate dielectrics\cite{HlinkaBCCD,MatoK2SeO4}. As a matter of fact,  soft polarization branches ensure a simultaneous {\it instability with respect to both the homogeneous and the staggered polarization}, and this seems to be the essential prerequisite of an AF material\cite{Dvorak,Sama70}.

Obviously, among the modes of the same irreducible representation, the temperature dependence can be shared due to the mode mixing. In particular, modes listed within the same column in Table I. are coupled in the AF phase. For example, we have verified that the overall IR plasma frequency\cite{Hlin06} of all $B_{3u}$ modes observed below 100\,cm$^{-1}$  is close to the IR plasma frequency of Last soft mode $\Omega_{\rm Last}=618$\,cm$^{-1}$ determined from the analysis of the cubic phase spectra\cite{Hlin06}.
 Since the bare $ X_5', \Sigma_1,$ and $S_1$ modes have the $B_{3u}$ symmetry but no intrinsic IR strength, the relative integral intensities of the $B_{3u}$ modes observed   in the conductivity spectrum below 100\,cm$^{-1}$ (Fig.\,5) can be directly interpreted as a measure of their eigenvector exchange with the pure Last mode.\cite{Hlin06} This mixing is visibly quite considerable.

 Are the observed temperature variations of phonon frequencies large or small ?
Within the Landau theory of the second-order structural phase transition,
 the squared soft-phonon frequency shows a linear temperature dependence (Cochran law). A stronger, {\it nonlinear} temperature dependence is expected below a first-order phase transition point,  but the relation
of the inverse static permittivity to the soft-mode frequency via Lyddane-Sachs-Teller relation should be still valid.
 In order to estimate the expected soft-mode frequency, we have thus multiplied the inverse of the static limit of the fitted permittivity  by the square of the  mode plasma frequency of the Last mode ($\Omega_{\rm Last}=618$\,cm$^{-1}$),
 and traced the resulting  temperature dependence of the squared bare Last mode frequency  ($\omega_{\rm Last, i}^2=\Omega_{\rm Last}^2 / \epsilon_i(0)$, $i=x,y,z$) in the  panel (h) of Fig.\,\ref{6a7}.
As it is clear from Fig.\,\ref{6a7}, such bare mode frequencies show very similar temperature dependence as most of the polar and nonpolar mode frequencies directly measured below 100\,cm$^{-1}$.

Let us stress that the $A_{g}$, $ B_{1g}$ and $B_{2g}$ Raman spectra show  additional soft modes, with  frequencies above
of 100\,cm$^{-1}$, and an even stronger temperature dependence (Fig.\,\ref{6a7}). These modes correspond well to the soft mode reported in Refs.\,\cite{Role89,Doba01}. It is natural to ascribe them to descendants of the $R_{25}$ rigid oxygen octahedra tilt mode, known  as the soft mode of the structural phase transition of SrTiO$_3$ crystal\cite{Cowl}. Indeed, the $R_{25}$ mode components associated with oxygen octahedra tilts around the $x,y$ and $z$ orthorhombic axes  do transform  precisely as the $A_{g}$, $ B_{1g}(y)$ and $B_{2g}(x)$ irreducible representations. A recent IXS study concluded that
 $R_{25}$ mode plays only a passive role of a triggered mode\cite{Taga13}, since its frequency shows virtually no temperature dependence in the cubic phase.
However, the mode investigated there had a frequency of about 50\,cm$^{-1}$ only, and according to our analysis, it was the Pb  ion vibration of $R_{15}$ symmetry. On the contrary, the present results indicate that the actual  $R_{25}$ oxygen octahedra tilt mode should have  a frequency of about 100-150 cm$^{-1}$ in the cubic phase.

Finally, panel (h) of Fig.\,\ref{6a7} reveals a remarkable anisotropy of the low-frequency permittivity  in the AF phase. Within the theory of Ref.\,\onlinecite{Taga13}, the  Curie-Weiss law for the AF phase was derived from  free-energy terms $\delta_{i} P_i^2 \rho^2$, $i=x,y,z$, where  $P_i$, are components of the macroscopic polarization, $\rho$ is the order parameter (staggered polarization) and $\delta_{x}=\delta_{y}$ and $\delta_{z}$ are positive coupling constants. The same terms determine also the magnitudes of AF coercive fields\cite{Taga13}. The anisotropy shown in panel (h) of Fig.\,\ref{6a7}  suggests  either that $\delta_{y} \approx 3\delta_{x}$ or that there is some other reason for preferential suppression of the  $\epsilon_{yy}$.

In summary, this polarized Raman, IR and THz spectroscopic study of  AF PbZrO$_3$ single crystals  established that there are several low frequency modes with  anomalously temperature dependent phonon frequencies distributed among all seven active irreducible representations.
We conclude that the modes around 130\,cm$^{-1}$ are associated with oxygen octahedra tilt vibrations, while those below about 100\,cm$^{-1}$ are due to the Pb ion fluctuations. Softening of the latter can be  understood as a consequence  of a soft and flat  phonon branch, without having to recall specific biquadratic couplings for each such mode separately. We argue that, in general, such soft and flat polarization fluctuation branches are expected to occur in AF materials.  We have also found that the low temperature dielectric tensor of PbZrO$_3$ is highly anisotropic and that this anisotropy originates from the anisotropic hardening of the Last mode components. We hope that the present systematic survey of low-frequency phonon modes will help in understanding and modeling of finite-temperature properties of PbZrO$_3$ and other antiferroelectric oxides.

\section{acknowledgments}
 Authors are indebted to J. Petzelt for critical reading of the manuscript.
This work was supported by the Czech Science Foundation (Project GACR 13-15110S).


\begin{thebibliography}{99}

\bibitem{Rabe13} K. M. Rabe, in {\it Functional Metal Oxides: New Science and Novel Applications}, edited by Satish Ogale and V. Venkateshan (Wiley, Hoboken, NJ, 2013).


 \bibitem{LiuH11} H. Liu and B. Dkhil, Z. Kristallogr. {\bf 226}, 163 (2011).

\bibitem{Misc06} A. S. Mischenko, Q. Zhang, J. F. Scott, R. W. Whatmore, and N. D. Mathur, Science {\bf 311}, 1270 (2006).
\bibitem{Taga13} A. K. Tagantsev,	
 K. Vaideeswaran,	 S. B. Vakhrushev,	 A. V. Filimonov,	 R. G. Burkovsky,	 A. Shaganov,	 D. Andronikova,	 A. I. Rudskoy,	A. Q. R. Baron,	 H. Uchiyama,	 D. Chernyshov,	 A. Bosak,	 Z. Ujma,	 K. Roleder,	A. Majchrowski,	 J.-H. Ko and N. Setter, 	Nature communications {\bf 4} 2229 (2013).

 \bibitem{Reye13} S.E.  Reyes-Lillo and K. M. Rabe, Phys. Rev. B {\bf 88}, 180102 (2013).

  \bibitem{Lines} M. E. Lines, A. M. Glass, Principles and Applications of Ferroelectrics and Related Materials (Oxford University Press, 1977).
\bibitem{Shir50} G. Shirane, E. Sawaguchi, and A. Takeda, Phys. Rev. {\bf 80}, 482 (1950).

\bibitem{Bhal00} A.S. Bhala, R. Guo, and R.Roy, Mat. Res. Innovat. {\bf 4}, 3 (2000).

\bibitem{Fuji68} H. Fujishita and S. Hoshino, J. Phys. Soc. Jpn. {\bf 53}, 273 (1968).

\bibitem{Coch68} W. Cochran and A. Zia, Phys. Stat. Sol. {\bf 25}, 273 (1968).

\bibitem{Fuji84} H. Fujishita and S. Hoshino, J. Phys. Soc. Jpn. {\bf 53}, 226 (1984).

\bibitem{Wagh97} U. V. Waghmare and K. M. Rabe, Ferroelectrics {\bf 194}, 135 (1997).

\bibitem{Hola73} J. Holakovsky, Phys. Stat. Sol. (b) {\bf 56}, 615 (1973).

\bibitem{Doba01} P. S. Dobal, R. S. Katiyar, S. S. N. Bharadwaja, and S. B. Krupanidhi, Appl. Phys. Lett. {\bf 78}, 1730 (2001).


\bibitem{Role89} K. Roleder, G. E. Kugel, M. D. Fontana, J. Handerek, S. Lahlou, and C. Carabatos-Nedelec, J. Phys.: Condens. Matter {\bf 1}, 2257 (1989).

\bibitem{Osta01} T. Ostapchuk, J. Petzelt, V. Zelezny, S. Kamba, V. Bovtun, V. Porokhonskyy, A. Pashkin, P. Kuzel, M. D. Glinchuk,
I. P. Bykov,   B. Gorshunov,  and M. Dressel, J. Phys. Cond. Matter. {\bf 13}, 2677 (2001).


\bibitem{Ghos99} Ph. Ghosez, E. Cockayne, U. V. Waghmare, and K. M. Rabe, Phys. Rev. B {\bf60}, 836 (1999).

\bibitem{Cock00} E. Cockayne and K. M. Rabe, J. Phys. Chem. Solids {\bf 61}, 305 (2000).


\bibitem{Leun02} K. Leung, E. Cockayne, and A.F. Wright, Phys. Rev. B {\bf 65}, 214111 (2002).

\bibitem{Leun03} K. Leung, Phys. Rev. B {\bf 67}, 104108 (2003).


\bibitem{Shir70} G. Shirane, J. D. Axe, J. Harada, and J. P. Remeika, Phys. Rev. B {\bf 2}, 155 (1970).
\bibitem{Tome12} I. Tomeno, J. A. Fernandez-Baca, K. J. Marty, K. Oka, and Y. Tsunoda, Phys. Rev. B {\bf 86}, 134306 (2012).
 \bibitem{Kemp06} M. Kempa, J. Hlinka, J. Kulda, P. Bourges, A. Kania, and J. Petzelt, Phase Transitions {\bf 79}, 351 (2006).
  \bibitem{Hlin11} J. Hlinka, P. Ondrejkovic, M. Kempa, E. Borissenko, M. Krisch, X. Long, and Z.-G. Ye, Phys. Rev. B {\bf 83}, 140101 (2011).
   \bibitem{Cowl}  R. A. Cowley, Phys. Rev. {\bf 134}, A981 (1964).

 \bibitem{KoJH13} J.-H. Ko, M. G\'{o}rny, A. Majchrowski, K. Roleder, and A. Bussmann-Holder, Phys. Rev. B 87, 184110 (2013).
\bibitem{Buix11}  E. Buixaderas, D. Nuzhnyy, J. Petzelt, L. Jin, and D. Damjanovic, Phys. Rev. B 84, 184302 (2011).



\bibitem{Dec89} J. Dec and J. Kwapulinski, J. Phys. Condens. Matter {\bf 1}, 3389 (1989).

\bibitem{Raman1} J. Hlinka, I. Gregora, J. Pokorny, C. H\'{e}rold, N. Emery, J. F. Marech\'{e}, and P. Lagrange, Phys. Rev. B 76, 144512 (2007).
\bibitem{Raman2} F. Borodavka, I. Gregora, A. Bartasyte, S. Margueron, V. Plausinaitiene, A. Abrutis, and J. Hlinka, J. Appl. Phys. {\bf 113}, 187216 (2013).

\bibitem{Hlin08} J. Hlinka, T. Ostapchuk, D. Nuzhnyy, J. Petzelt, P. Kuzel, C. Kadlec, P. Vanek, I. Ponomareva, and L. Bellaiche, Phys. Rev. Lett. {\bf 101}, 167402 (2008).

 \bibitem{DielExp} D. Nuzhnyy, J. Petzelt, M. Savinov, T. Ostapchuk, V. Bovtun, M. Kempa, J. Hlinka, V. Buscaglia, M.T. Buscaglia, and P. Nanni, Phys. Rev. B 86, 014106 (2012).

\bibitem{HlinkaBCCD} J. Hlinka, M. Quilichini, R. Currat, and J.-F. Legrand, J. Phys.: Condens. Matter {\bf 8}, 8207 (1996); {\bf 8}, 8221 (1996).
\bibitem{MatoK2SeO4} I. Etxebarria, M. Quilichini, J.M. Perez-Mato, P. Boutrouille, F.J. Zuniga, and T. Breczewski, J. Phys. Condens. Matter {\bf 4}, 8551 (1992).
\bibitem{Dvorak} V. Dvorak, Phys. Status Solidi {\bf 14}, K161 (1966).
\bibitem{Sama70} G. A. Samara, Phys. Rev. B 1, 3777 (1970).
\bibitem{Hlin06} J. Hlinka, J. Petzelt, S. Kamba, D. Noujni, and T. Ostapchuk, Phase Transit. {\bf 79}, 41 (2006).





\end{thebibliography}
\end{document}